\documentclass[preprint2]{aastex}

\shorttitle{MF and the rotation curve of M31}
\shortauthors{Ruiz-Granados et al.}

\begin{document}

\title{Magnetic fields and the outer rotation curve of M31}

\author{B. Ruiz-Granados and J.A. Rubi\~{n}o-Mart\'{i}n}
\affil{Instituto de Astrof\'{i}sica de Canarias (IAC), 
E-38200, La Laguna, Tenerife (Spain)
\and Departamento de Astrof\'{i}sica, Universidad de La Laguna, E-38205, La Laguna, Tenerife (Spain)  
}
\authoraddr{C/V\'{i}a L\'{a}ctea, s/n,
E-38200, La Laguna, Tenerife (Spain) \\}
\author{E. Florido and E. Battaner}
\affil{Departamento F\'{i}sica Te\'{o}rica y del Cosmos. Universidad de Granada, Granada(Spain)  
\and Instituto de F\'{i}sica Te\'{o}rica y Computacional Carlos I, Granada(Spain)}


\begin{abstract}
  Recent observations of the rotation curve of M31 show a rise
  of the outer part that can not be understood in terms of standard
  dark matter models or perturbations of the galactic disc by M31's
  satellites. Here, we propose an explanation of this dynamical
  feature based on the influence of the magnetic field within the thin
  disc.
  We have considered standard mass models for the luminous mass
  distribution, a NFW model to describe the dark halo, and we have
  added up the contribution to the rotation curve of a magnetic field
  in the disc, which is described by an axisymmetric pattern.
  Our conclusion is that a significant improvement of the fit in the
  outer part is obtained when magnetic effects are considered. The
  best-fit solution requires an amplitude of $\sim 4$~$\mu$G with a
  weak radial dependence between 10 and 38~kpc.
\end{abstract}


\section{Introduction}
Recent high sensitivity measurements of the rotation curve of
M31~\citep[][hereafter C09 and C10 respectively]{chemin09,corbelli10}
suggest that it highly rises at the outermost part of the disc of
M31 ($r \ga 25-30$~kpc) (see Figure~\ref{fig1}). This behaviour cannot be considered an exception.
Similar outer
rising rotation curves can be found in other galaxies. \cite{noordermeer07}
stated that ``in some cases, such as UGC 2953, UGC 3993 or UGC 11670 there are
indications that the rotation curves start to rise again at the outer edges of
the HI discs'', suggesting follow-up observations of higher sensitivity to
investigate this fact in more detail.
Other examples could be found in some rotation curves provided
by~\cite {spano08}, in particular the curves of UGC 6537 (SAB(r)c),
UGC 7699 (SBcd/LSB), UGC 11707 (Sadm/LSB) or UGC 11914 (SA(r)ab)
measured by means of Fabry-Perot spectroscopy. Other potential
examples could be NGC 1832 and NGC 2841~\citep{kassin06}, even with a
larger error, and ESO 576-G51 and ESO 583-G7~\citep{seigar05}, even if
the outer rising is small. Potential nearby candidates could be NGC 3198, DDO 154 or NGC 7331~\citep{deblok08}.
Therefore, outer rising rotation curves are not uncommon in spirals,
being the case of M31 the most representative. This puzzling dynamic
feature requires a theoretical explanation.
A detailed study by C09 showed that the NFW, Einasto or
pseudo-isothermal dark matter halos fail to reproduce the exact shape
of the rotation curve of M31 in the outer region and found no
differences between the various halo shapes. Moreover, they found new
HI structures as an external arm and thin HI spurs in the outskirts of
the disc. These spurs have been also observed by~\cite{braun09}. 
Thus, a primarily explanation of this gas perturbations could be the
interactions with the M31's satellites as NGC
205~\citep{geehan06,corbelli10}. The main problem of matching the properties of
the giant stellar stream observed in the south of M31~\citep{ibata01}
is that the orbital of the companion that produce the stream is not
constrained satisfactorily~\citep{fardal06} or high velocities for the
radial orbits are found~\citep[see e.g.][]{howley08}. Although the northeastern (NE) and southwestern (SW) parts of M31 show different kinematical properties, both suggest a rise in the outer part of the rotation curve (see Figure 5 and 6 in C09 and C10, respectively). 

In this work, we show that magnetic dynamic effects constitute a clear
and simple basis to interpret this feature. Some particular models of
magnetically driven rotation curves have been
presented~\citep{nelson88,battaner92,battaner95,battaner00,battaner07,kutschera04,tsiklauri08}.
Although those models were originally developed to explain flat
rotation curves without dark matter, our purpose here is more
conventional and we will consider the contribution of both a NFW dark
matter halo and a magnetic field added to match the velocity in the
outermost region.

Magnetic fields are known to slowly decrease with radius~\citep[see
  e.g.][]{beck96,han98,beck01,beck04,beck05}, and therefore they
become increasingly important at the rim. For large radii, an
asymptotic $1/r$-profile for the field strength provides an asymptotic
vanishing magnetic force. This $1/r$-profile has been found to match the polarized synchrotron emission of the Milky Way~\citep{ruiz-granados10} and NGC 6946~\citep{beck07}, and will be considered in this work.
%

\section{Mass models and the magnetic field of M31}
\label{sec:massmodels}

In this section, we briefly present the luminous and dark mass models
used to describe the observational rotation curve of M31. As we are
interested in the outer region (i.e. distances higher than 10 kpc), we
follow C10 and we neglect the bulge contribution.
We will also adopt from C09 and C10 the parameters describing stellar
disc and gaseous distribution, respectively.
Therefore, we will only leave as free parameters those describing two
physical components: the dark matter halo and the magnetic field.

\subsection{Disc model}

The stellar disc is assumed to be exponential~\citep{freeman70}, being
the surface mass density
\begin{equation}
\label{eq:surface_density}
\Sigma = \Sigma_{0}\exp{\left ( \frac{-r}{R_{d}} \right )},
\end{equation}
where $\Sigma_{0}$ is the central density of stars and $R_{d}$, the
radial scale factor.

The contribution to the circular velocity of the stellar disc
is~\citep[see][]{binney87}
\begin{equation}
\label{eq:vel_disc}
v_{d}(r) = 4 \pi G \Sigma_{0} R_{d} y^{2} [I_{0}(y)K_{0}(y) - I_{1}(y)K_{1}(y)],
\end{equation}
where $y = r/2R_{d}$ and $I_{0},K_{0},I_{1},K_{1}$ are the modified Bessel's
functions for first and second order.

%
The gaseous disc mainly contains HI, molecular gas and helium. The
estimated gaseous distribution of M31 represents approximately a 9\%
(for C09) and 10\% (for C10) of the total mass of the stellar disc
(see Table~\ref{tab1}).

For the spatial distribution of the gas, we again assume the same
exponential law given by Equation~(\ref{eq:surface_density}).

\subsection{Halo model}
\label{sec:halo}
The dark matter halo is described here by a NFW profile~\citep{navarro96}
\begin{equation}
\label{eq:density_nfw}
\rho_{h}(r) =\frac{\delta_{c}\rho_{c}}{\frac{r}{R_{h}} \left [ 1 + \left ( \frac{r}{R_{h}} \right ) \right ]^{2}},
\end{equation}
where $\delta_{c}$ is a characteristic density contrast, $\rho_{c} = 3
H_{0}^{2}/ 8 \pi G$ is the critical density and $R_{h}$ is the radial
scale factor.
The contribution to the circular velocity due to this profile is given
by~\citep{navarro97}
\begin{equation}
v_{h}(x) = V_{200}\sqrt{\frac{1}{x}\frac{\ln(1+cx) - \frac{cx}{1+cx}}{\ln(1+c)-\frac{c}{1+c}}},
\end{equation}
where $V_{200}$ is the velocity at the virial radius $R_{200}$ which
is assumed $\sim 160$ kpc for C09 and $R_{200} \sim 200$~kpc for C10,
$c$ is the concentration parameter of the halo which is defined as $c
= R_{200} / R_{h}$ and $x$ is $r / R_{200}$. According to C09 and C10,
the total dark halo mass at the respective virial radius is $M_{200}
\sim 10^{12}$ $M_{\odot}$.
Our free parameters are $V_{200}$ and $c$. Values from C09 are $c =
20.1\pm2.0$ and $V_{200} = 146.2 \pm 3.9$ $km/s$. Values from C10 are
$c = 12$ and $V_{200} \sim 140$ $km/s$.

\subsection{The regular magnetic field of M31}
The first measurements of the magnetic field of M31 were obtained by
\cite{beck82} using the polarized synchrotron emission at 2700 MHz, showing that a magnetic pattern was aligned with HI structures and formed a ring at $r \approx 10$~kpc. They obtained a strength of $B_{reg} = 4.5 \pm{1.2}$ $\mu$G. \cite{berkhuijsen03} used observations of polarized emission to show that M31 hosts an axisymmetric field. \cite{han98} used Faraday rotation measurements to show that the regular field could extend at least up to galactocentric distances of $r \sim 25$~kpc without significant decrease of the strength and at least with a $z \sim 1$ kpc of height above the galactic plane. However, recently,~\cite{stepanov08} have shown that these results contained a significant contribution of Galactic foregrounds and so, it is difficult to infer the model from these measurements. In any case, as we show below, the detailed structure of the field is not relevant for the rotation curve.

\cite{fletcher04} presented a detailed study of the regular field of
M31 based on multi-wavelength polarized radio observations in the
region between 6 and 14 kpc. By fitting the observed azimuthal
distribution of polarization angles, they found that the regular
magnetic field follows an axisymmetric pattern in the radial range
from 8 to 14~kpc. The pitch angle decreases with the radial distance,
being $p \sim -16^{\circ}$ for distances $r < 8$~kpc, and $p \sim
-7^{\circ}$ for $r < 14$ kpc. This fact implies that the field becomes
more tightly wound with increasing galactocentric distance. They found
a total field strength (i.e. regular and turbulent components) of
$\approx 5$ $\mu$G. For the regular field they found that it became
slowly lower, reaching at $r \sim 14$~kpc a strength of $\sim 4.6$
$\mu$G.

%
In this paper, our basic assumption is that the regular magnetic field
of M31 is described by an axisymmetric model, that extends up to 38~kpc. For this model, the components in cylindrical coordinates are given by
\begin{eqnarray}
  \label{eq:ASS}
    \label{eq:ASS_r}
    B_{r} &= B_{0}(r) \sin(p) \\
    \label{eq:ASS_phi}
    B_{\phi} &= B_{0}(r) \cos(p), 
\end{eqnarray}
where $p$ is the pitch angle and $B_{0}(r)$ is the field strength as a function
of the radial distance.
As shown below, from the point of view of the description of the rotation curve,
the relevant component is $B_{\phi}$.
Here, we will assume that $p = 0^{\circ}$, as we are mostly interested
in the outer region where $p$ is very low. Several previous probes
also indicates low values of $p$. In any case, this is not a strong
assumption in the sense that a non-zero $p$ value can be absorbed into
the field strength ($B_{1}$) as a different amplitude.
For the field strength, as a baseline computation, we shall consider a
radial dependence of $B_0(r)$ or equivalently $B_{\phi}$ given by
\begin{equation}
B_{\phi}(r)=\frac{B_{1}}{1+\frac{r}{r_{1}}},
\label{eq:radial}
\end{equation}
where $r_1$ represents the characteristic scale at which $B_{0}(r)$
decreases to half its value at the galactic centre and $B_{1}$ is an
amplitude in which we are absorbing the $\cos(p)$ factor.
This expression has an appropriate asymptotic behaviour, in the sense
that we obtain a finite value when $r$ is close to the galactic center
($r \rightarrow 0$), and asymptotically tends to $\propto 1/r$ when $r
\rightarrow \infty$, as suggested \cite{battaner07}.
Observations carried out by \cite{fletcher04} established that
\begin{equation}
\label{eq:prior}
B_{\phi}(r = 14 kpc) \approx 4.6 \mu G. 
\end{equation}
This observational value at this radius was
considered as fixed in our baseline computation. By
substituting into Equation~(\ref{eq:radial}), we can find a relation between
$B_{1}$ and $r_{1}$ that allows us to re-write Equation~(\ref{eq:radial}) in
terms of a single free-parameter ($r_{1}$)
\begin{equation}
B_{\phi}(r) = \frac{4.6 r_{1} + 64.4}{r_{1}+r},
\label{eq:radial2}
\end{equation}
where $r$ is given in kpc and $B_{\phi}(r)$ in $\mu$G.
%

\section{Methodology}

\subsection{Observational rotation curve of M31}
We have considered the two datasets from C09 and C10. They consist on
a set of 98 and 29 measurements of the circular velocity (and their
associated error bars) respectively, which were obtained with the
high-resolution observations performed with the Synthesis Telescope
and the 26-m antenna at the Dominion Radio Astrophysical
Observatory~(C09) and with the wide-field and high-resolution HI
mosaic survey done with the help of the Westerbork Synthesis
Radiotelescope and the Robert C. Byrd Green Bank Telescope
(GBT)~\citep{braun09}. For our purposes, we consider only distances
higher than $r > 10$~kpc to illustrate the effects of the magnetic
field.
The actual data points on the rotation curve were obtained after
fitting a tilted ring model to the data, and assuming a distance to
M31 of 785~kpc. C09 derived a value of the inclination angle of $i
\sim 74^{\circ}$, which is lower than that derived from optical
surface photometry measurements~\citep{walterbos87} and by C10 ($i
\sim 77^{\circ}$).
The data points are plotted in Figure~\ref{fig1}, in two separate
panels.

\subsection{Influence of the magnetic field on the gas distribution}
The presence of a regular magnetic field affects the gas
distribution~\citep{battaner92,battaner95,battaner00}.
The fluid motion equation can be written as~\citep[see
  e.g.][]{battaner96}
\begin{equation}
\label{eq:plasma_movimiento_mag}
\rho \frac{\partial \vec{v}_{0}}{\partial t} + \rho \vec{v}_{0} \cdot \vec{\nabla}\vec{v}_{0} + \vec{\nabla} P = n \vec{F} + \frac{1}{4 \pi} \vec{B}\cdot \vec{\nabla} \vec{B} - \nabla \left(  \frac{B^{2}}{8 \pi} \right),
\end{equation}
where $\rho$ is the gas density; $\vec{v}_{0}$, the velocity of the
fluid; $P$, the pressure; $\vec{F}$, the total force due to gravity
and $\vec{B}$, the magnetic field.
We assume standard MHD conditions i.e. infinite conductivity.
Equation~(\ref{eq:plasma_movimiento_mag}) is simplified by assuming
axisymmetry and assuming
pure rotation, where $\vec{v_{0}} = (v_{0r}, v_{0\phi}, v_{0z}) = (0,
\theta, 0)$ even if these conditions are not necessary regarding the dynamic
effects in the radial direction. Taking into account all these facts, the motion equation
in the radial cylindrical coordinate is
\begin{equation}
\label{eq:plasma_movimiento_cil}
\rho\left ( -\frac{d \Phi(r)}{d r}  + \frac{\theta^{2}}{r} \right ) - \frac{d P}{d r} - F_{r}^{mag} = 0,
\end{equation}
where $\Phi(r)$ is the gravitational potential; $F_{r}^{mag}$, the
radial component of the magnetic force, and $P$ the pressure of the
fluid. We can assume that pressure gradients in the radial direction
are negligible~\citep{battaner00}. In this case, then the radial
component of the magnetic force is given by
\begin{equation}
\label{eq:fuerza_magnetica_r}
F_{r}^{mag} = \frac{1}{4 \pi} \left( \frac{B_{\phi}^{2}}{r} + \frac{1}{2}\frac{d B_{\phi}^{2}}{d r}  \right ),  
\end{equation}
and the contribution of the magnetic field to the circular velocity is
given by
\begin{equation}
\label{eq:vel_circ_mag}
 v_{mag}^{2} = \frac{r}{4 \pi \rho} \left( \frac{B_{\phi}^{2}}{r} +
 \frac{1}{2}\frac{d B_{\phi}^{2}}{d r} \right ).
\end{equation}
%

\subsection{Modelling the rotation curve}

The rotation curve is obtained, as usual, by quadratic summation of
the different contributions
\begin{equation}
\label{eq:vel_rot_total}
  \theta(r)^{2} = v_{b}(r)^{2} + v_{d}(r)^{2} + v_{h}(r)^{2} +
  v_{mag}(r)^{2},
\end{equation}
where we explicitly set $v_{b}(r) = 0$ as mentioned above.
\subsection{Model selection}
For the luminous mass models, the different parameters are considered
as fixed values in our analysis (see Table~\ref{tab1}).
For the NFW dark halo, we constrain the $V_{200}$ and $c$ parameters,
allowing a range for $V_{200}$ between $100$ and $220$ km/s with steps
of 0.5 km/s and $c \in [5,30]$ with steps of 0.3.
The contribution of the magnetic field to the rotation curve is fitted
through one free parameter, $r_{1}$, that we are considering which is
equivalent to fit $B_{\phi}$ as we discussed above. For this
parameter, we have explored values in the range from 1 to 1000~kpc.
Our analysis is based on a reduced-$\chi^{2}$ as the goodness-of-fit
statistic. Thus, the best-fit parameters are obtained by minimizing
this function
\begin{equation}
\chi^{2} = \frac{1}{N-M} \sum_{i=1}^{n}{\frac{(\theta^{obs}_{i} -
    \theta^{model}_{i})^{2}}{\sigma_{i}^{2}}},
\end{equation} 
where $N$ is the total number of points to which we have measured the
rotational velocity and depends on the considered dataset ($N = 74$
for C09 and $N = 27$ for C10) and $M$ is the number of free
parameters. The sum runs over the observational data points, being
$\theta^{obs}_{i}$ the observed velocity and $\theta^{model}_{i}$ the
modelled velocity, which depends on the particular model. We shall
consider two models: one without magnetic contribution (DM) and
another with the magnetic field (DM+MAG). Finally, $\sigma_{i}$ is the
observational error bar associated to each data point.
%

\begin{table*}
\vspace{0.2cm}
\begin{center}
\begin{tabular}{@{} l c r}
\hline \hline
Dataset    &    Disc parameters  \\ 
\hline \hline
C09        &     $R_{d} = 5.6$~kpc   \\
           &     $M_{d} = 7.1 \times 10^{10}$ $M_{\odot}$  \\
           &     $M_{gas} \sim 6.6 \times 10^{9}$ $M_{\odot}$\\ \hline

C10        &     $R_{d} = 4.5$~kpc   \\
           &     $M_{d} = 8.0 \times 10^{10}$ $M_{\odot}$  \\
           &     $M_{gas} \sim 7.7 \times 10^{9}$ $M_{\odot}$\\ \hline   
\hline
\end{tabular}
\end{center}
\tablenum{1}
\caption{
Fixed parameters for bulge and disc.\label{tab1}}
\end{table*}


\section{Results and discussion}
\label{sec:results}
Our main results are summarized in Figure~\ref{fig1}. The dotted line
shows our best-fit rotation curve for C09 and C10 when considering
only the usual dynamical components (the stellar component and the
dark halo at this range of distances), while the solid line shows the
result when adding up also the magnetic contribution.

The most important result is that, as shown in \cite{battaner00},
the effects of magnetism on the rotation curve are only relevant at
large radii (in this case of M31, at distances larger than about
25~kpc).
In this sense, the radial range of the datasets of C09 and C10 are
optimal to observe the magnetic effects.

Table~\ref{tab2} summarizes our results for the best-fit solutions
without (labelled as DM model) and with magnetic field influence (DM +
MAG model).
%
\begin{table}
\begin{center}
\begin{tabular}{@{} l c c c r}
\hline
\hline
Model      &  Parameters          &       C09                &  C10  \\ 
\hline
\hline 
DM         &    $V_{200}$ (km/s)   &      $160.2\pm{2.0}$         &     $132.1^{+5.7}_{-5.4}$   \\
           &    $c$               &      $12.3\pm{0.6}$         &      $19.1^{+2.4}_{-2.2}  $   \\
           &    $\chi^{2}$         &       19.8                   &       1.1      \\ \hline

DM + MAG   &    $V_{200}$ (km/s)   &       $ 133.8^{+1.7}_{-1.3}$    &      $120.0^{+4.7}_{-4.0}$  \\
           &    $c$                &      $ 22.7^{+1.2}_{-1.1}$     &      $25.0^{+2.8}_{-2.9}$    \\
           &    $r_{1}$ (kpc)      &       $> 888.0 $             &      $> 185.0$    \\
           &    $\chi^{2}$         &        13.3                  &      0.6      \\
    
\hline \hline
\end{tabular}
\end{center}
\tablenum{2}
\caption{ Best-fit for the rotation curve with and without the
  contribution of magnetic fields for $r \ga 10$ kpc.\label{tab2}}
\end{table}
As shown, magnetic effects on the gaseous disc significantly decrease
the value of the reduced $\chi^{2}$ statistic for both
datasets. Specially for C09, the fit is significantly improved when
taking into account a new parameter ($\Delta \chi^2 = 6.5$).
The radial scale factor of the magnetic field ($r_1$) is unconstrained
in both cases, but shows clear preference for high values, which means
that the best-fit solution for the field slowly decreases with the
radial distance in the considered interval (i.e. between 10 and
38~kpc). 
For example, at $r \sim 38$~kpc, the field is found to be $B_{\phi}
\ga 4.4$ $\mu$G for C09 and $B_{\phi} \ga 4.0$ $\mu$G for C10 for this
best-fit solution. Both values are compatible with the strength of the field obtained by \cite{fletcher04} who found a nearly constant strength of the regular field of about $\sim 5\mu$G between 6~kpc and 14~kpc.
Moreover, when no radial variation of the strength is considered,
(i.e. if $r_{1} \rightarrow \infty$), and we fit for the amplitude
$B_{\phi}$, we obtain that $B_{\phi} = 4.7^{+0.6}_{-0.7}$ $\mu$G which
it is again compatible with results discussed above.
This suggests that the data do not require an important radial variation of the strength of the field for the considered range of distances, or in other words, the contribution of the second term in the r.h.s. of Equation~(\ref{eq:fuerza_magnetica_r}) is negligible (i.e. $dB_{\phi}^{2} / dr \ll B_{\phi}^{2} / r$). Therefore, if we had considered another radial profile (e.g. exponential), we would have found a large radial scale factor too. The azimuthal component of the field is practically constant between 10 and 38~kpc.

Our results imply large magnetic fields at large radii. How these are
produced lies beyond our scope. On the other hand, we would expect the extragalactic field to be also of
this order of magnitude. Theoretical predictions \citep{dar05} suggest
values of the level of few $\mu$G for the intergalactic magnetic fields (hereafter IGMF). \cite{kronberg94,govoni04,kronberg05} have reviewed observations of $\mu$G level in rich clusters, though no direct
measurements have been reported for the IGMF in the Local Group near M31.
The observational evidence for IGMF is still weak, but
quite strong IGMF near galaxies cannot be disregarded.

We finally note the apparent discrepancy between our derived
parameters for the DM model and those obtained by C09 and C10 (see
Sect.~\ref{sec:halo}). However, it is important to stress that we are
restricting the fit to the outer region of M31 ($r>10$~kpc). In this
outer range, there is a weaker dependence of the shape of the rotation
curve on the concentration parameter $c$, and thus lower values of $c$
are found because the fit tries to compensate the rising behaviour in the
outer part. 
The inclusion of the magnetic field corrects this apparent
discrepancy, and in this case the values of $c$ are now compatible
with those obtained by C09 and C10. 

\section{Conclusions}

The rotation curve of M31~\citep{chemin09,corbelli10} is rather
representative of the standard rotation curves of spirals for $r <
30$~kpc, but it seems to rise out to, at least 38~kpc, the
limiting distance of the observations. Indeed, this behaviour is not
restricted to M31, and we are probably dealing with a common dynamical
feature of many other spirals. Therefore, the outermost rising rotation curve is a very important theoretical challenge.

It is certainly a challenge as the standard dark matter halo models, in
particular the universal NFW profiles, do not account for this dynamical
unexpected behaviour.

A conventional galactic model, with bulge, disc and dark matter halo, has been shown
to provide good fit to the data in the range $r < 20$~kpc (C09, C10).
Here, we take advantage of these results and we do not fit any of the luminous components,
which are taken to be exactly the same as those proposed by C09 and C10.
We have restricted our study to the region $r > 10$~kpc, and we only fitted the parameters
describing the dark matter halo, and the magnetic field contribution.
Our main conclusion is that magnetic fields are not ignorable for explaining large-scale
dynamic phenomena in M31, producing a significant improvement of the fit of the rotation curve at
large distances. Moreover, the required field strength of the regular component ($B_{\phi} \sim 4$ $\mu$G) is fully consistent with the measured magnetic field in M31 at least up to $r \sim 15$~kpc.

This conclusion seems very reasonable, as magnetic fields are
amplified and act ``in situ'', and therefore they become increasingly
important at the rim, where gravity becomes weaker.

The best-fit model of the magnetic fields requires a field strength
slowly decreasing with radius. 
This slow decrease is compatible with
present values of the strength derived from observations of the polarized synchrotron emission of the disc, but clearly we need measurements of Faraday rotation of extragalactic sources at this large radii to confirm that the magnetic field is present up to this distance and to trace unambiguously the regular component. Hence, future experiments such as LOFAR\footnote{http://www.lofar.org} and SKA\footnote{http://www.skatelescope.org/}~\citep{beck09b}, will be extremely important, allowing a detailed explorations on the galactic edge as well as in the intergalactic medium.

\begin{figure}
\plotone{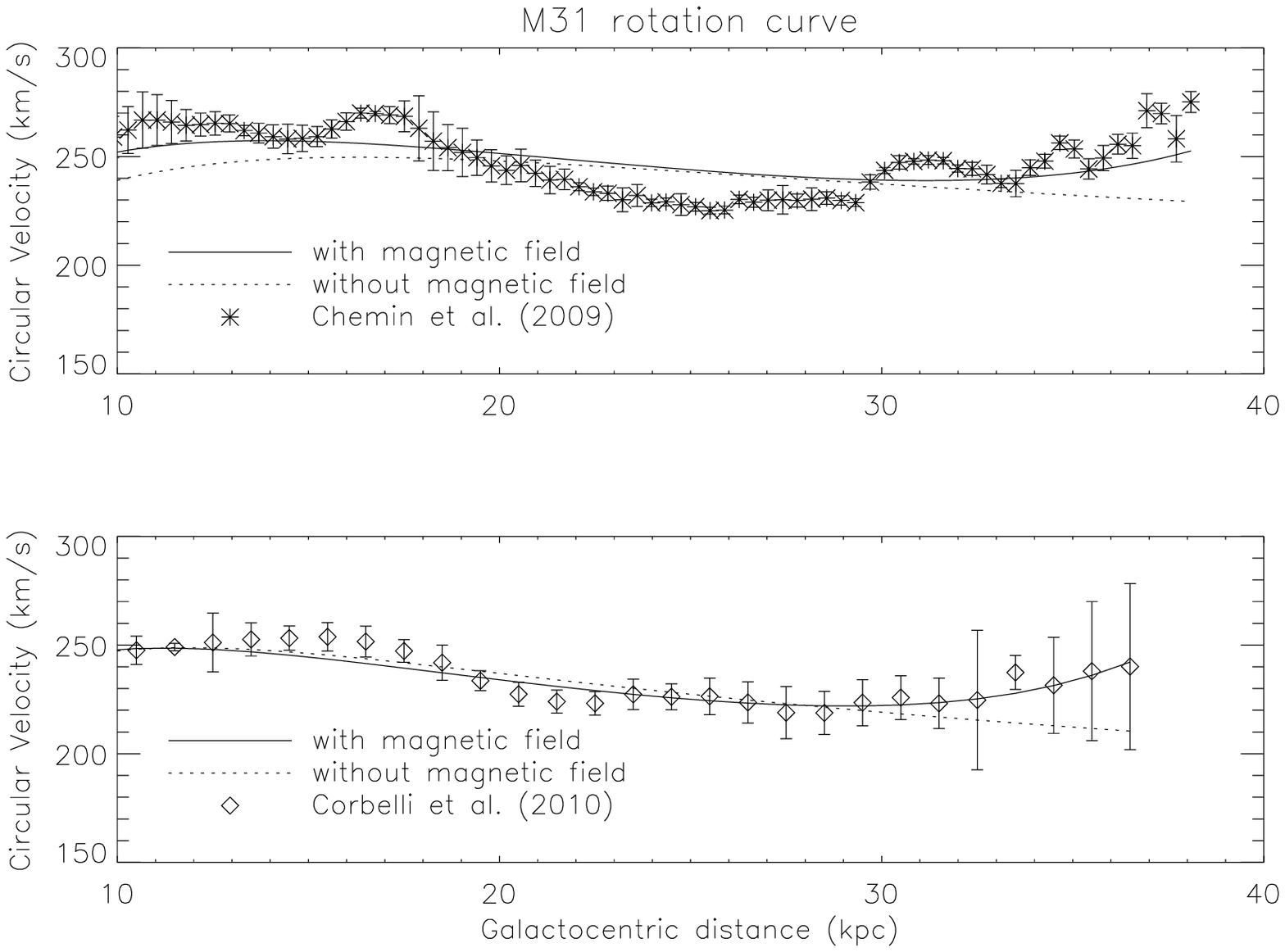}
\caption{
Best-fit solutions for the rotation curve of M31, with and without
  including the contribution of a regular magnetic field component. Top shows the C09 dataset and bottom the C10 dataset. Asterisks and rombhus represent the observational data with the associated error bars. The solid line is the best fit derived in this paper,
  including the contribution of the regular magnetic field over the gaseous disc. The dotted line is the best-fit model obtained without
  the contribution of the magnetic field.\label{fig1}}
\end{figure}

\placefigure{fig1}

%
\acknowledgements 
This work was partially supported by projects AYA2007-68058-C03-01 of
Spanish Ministry of Science and Innovation (MICINN), by Junta de
Andaluc\'ia Grant FQM108 and by Spanish MEC Grant AYA
2007-67625-C02-02. JAR-M is a Ram\'on y Cajal fellow of the MICINN.


\end{document}